\documentclass[12pt,
onecolumn,
amsmath,amssymb,
aps,
nofootinbib,
]{revtex4-2}

\usepackage[margin=1in]{geometry}
\usepackage{graphicx}
\usepackage{bm}
\usepackage{xcolor}
\usepackage{titlesec}
\usepackage{setspace}

\titlespacing*{\section}{0pt}{5pt}{3pt}

\begin{document}

\title{Thermodynamics as a tool for (quantum) gravitational dynamics}

\author{Ana Alonso-Serrano}
\email{ana.alonso.serrano@aei.mpg.de}
\affiliation{Max-Planck-Institut f\"ur Gravitationsphysik (Albert-Einstein-Institut), \\Am M\"{u}hlenberg 1, 14476 Potsdam, Germany}
\author{Marek Li\v{s}ka,\footnote{Corresponding author}}

\email{liska.mk@seznam.cz}
\affiliation{Institute of Theoretical Physics, Faculty of Mathematics and Physics, Charles University,
\\V Hole\v{s}ovi\v{c}k\'{a}ch 2, 180 00 Prague 8, Czech Republic}

\date{\today}

\begin{abstract}
The thermodynamics of local causal horizons has been shown to imply gravitational dynamics. In this essay, we discuss the principles underlying this observation, and its significance in our understanding of (quantum) gravity. We also show why the local thermodynamic methods cannot by themselves recover general relativity. Instead, they lead to the so-called Weyl transverse gravity. Because of this, local thermodynamic approaches avoid huge vacuum energy contributions to the cosmological constant. They even suggest a possible source for its small observed value. We also outline a way in which thermodynamics allows us to study low energy quantum gravitational effects. We arrive at quantum corrections to the gravitational equations which are suppressed by the Planck length squared.
\end{abstract}

\maketitle

\vspace{1.5 cm}

Essay written for the Gravity Research Foundation 2023 Awards for Essays on Gravitation

\newpage

\section*{Introduction}

The connection between black hole physics and thermodynamics was first established in the early seventies~\cite{Bekenstein:1973,Bardeen:1973,Hawking:1975}.
It relies on the interplay of  gravitational dynamics and quantum physics. The former yields a description of black hole mechanics analogous to the laws of thermodynamics, the latter then allows to identify black hole temperature~\cite{Hawking:1975,Visser:2003} and entropy~\cite{Bardeen:1973,Gibbons:1977,Wald:1993}. Due to this position at the intersection of gravitational and quantum paradigms, thermodynamics is often considered as a window into the nature of the, as of yet elusive, quantum theory of gravity~\cite{Wald:2001,Wuthrich:2019}.

We will argue that thermodynamics actually provides clues both about the classical structure of gravity, and about the low energy quantum gravitational effects. Moreover, it does so independently of any strong assumptions regarding the nature of quantum gravity. However, to do so, we must reverse the reasoning outlined in the previous paragraph. Rather than studying thermodynamic properties of black holes for a given theory of (quantum) gravity, we employ thermodynamic considerations to derive the equations governing gravitational dynamics. This was first realized by showing that Einstein equations are equivalent to thermodynamic equilibrium conditions imposed on local, observer-dependent causal horizons constructed at every spacetime point~\cite{Jacobson:1995ab}. Notably, while the resulting equations are classical, their derivation requires quantum physics in order to define the temperature of local causal horizons. Following that seminal paper, thermodynamic derivations have been extended to accommodate non-equilibrium entropy production~\cite{Eling:2006,Chirco:2010}, to derive equations of motion for modified theories of gravity~\cite{Jacobson:2012,Bueno:2017,Svesko:2017,Svesko:2019}, and were even rephrased as a variational principle~\cite{Padmanabhan:2010}.

Moreover, some approaches moved from thermodynamics to a notion of equilibrium stated completely in terms of quantum entanglement, both in the context of AdS/CFT correspondence~\cite{Lashkari:2014,Faulkner:2014,Faulkner:2017}, and independently of it~\cite{Jacobson:2016}. Notably, the entanglement equilibrium conditions imply semiclassical equations for gravitational dynamics in which the classical gravitational field is sourced by the quantum expectation value of the energy-momentum tensor. Thus, the resulting dynamics already includes quantum corrections, albeit only on the matter side.

Herein, we present a broad perspective on a new line of research we have been pursuing in the recent years. Our aim is to argue that thermodynamic/entanglement equilibrium methods may be pushed even further, to provide nontrivial insights into the nature of quantum gravity. Our conclusions apply to any method for deriving gravitational dynamics from local equilibrium conditions\footnote{We do not discuss proposals based on equilibrium conditions applied to globally defined surfaces~\cite{Akbar:2007,Verlinde:2011}. While related to the local approaches, they nevertheless have several distinct features and, in general, our conclusions do not apply to them.}. Hence, rather than focusing on one particular method, we instead provide a general argument valid for any local equilibrium approach.

\section*{Minimal requirements for local equilibrium approaches}

To rely on local equilibrium conditions as a tool for studying gravitational dynamics, it is crucial to understand both explicit and implicit assumptions involved in using them. Hence, we begin by briefly reviewing the necessary ingredients to connect local equilibrium conditions with gravity.

First, one needs to define a notion of temperature of a local causal horizon. This is done by invoking the Unruh effect, a standard result of flat spacetime quantum field theory. To make the Unruh effect locally valid even in a generic curved spacetime, the local equilibrium approaches typically assume the Einstein equivalence principle: ``Fundamental non-gravitational test physics is not affected, locally and at any point of spacetime, by the presence of a gravitational field~\cite{Casola:2015}'', which of course played a pivotal role also in the original development of general relativity. The Einstein equivalence principle can be shown to imply that gravity is described by the metric (and possibly additional fields), but does not constrain the gravitational dynamics any further~\cite{Casola:2015} (unlike the strong equivalence principle~\cite{Casola:2015,Casola:2014}, which we do not need for our purposes~\cite{Chirco:2010,Alonso:2021}). Hence, to introduce the Unruh temperature, we already constrain the kinematic description of gravity, but leave its dynamics arbitrary. At the level of classical physics, the Einstein equivalence principle is uncontroversial. However, its status when quantum effects become important remains unclear~\cite{Majhi:2023}. Luckily, the equivalence principle is only really needed to invoke the Unruh effect. Hence, we can proceed with a weaker assumption that, if the spacetime curvature effects can be disregarded, uniformly accelerating observers perceive a thermal bath of particles. In fact, this statement has been proposed as a suitable quantum reformulation of the equivalence principle~\cite{Smolin:2016}.

Second, we require that entropy of a locally constructed observer-dependent causal horizon\footnote{To be precise, we should talk about entropy of closed causal horizons, realized, e.g. as causal diamonds or stretched light cones. The infinite extension of the usual acceleration horizons makes defining the associated entropy more problematic~\cite{Carroll:2016,Svesko:2017} However, we omit this rather technical distinction in the present text for the sake of clarity.} is given by the same expression as that of a black hole's horizon~\cite{Jacobson:2003}. While the temperature only concerned the kinematic description of gravity, the precise form of the entropy determines the gravitational dynamics~\cite{Eling:2006,Padmanabhan:2010,Jacobson:2012,Svesko:2017,Svesko:2019}. In particular, to derive Einstein equations, we require that entropy of both local causal horizons and black hole horizons is proportional to their area $\mathcal{A}$ and given by the Bekenstein expression, $S=k_{\text{B}}\mathcal{A}/\left(4l_{\text{P}}^2\right)$~\cite{Jacobson:1995ab,Jacobson:2016}. In other words, black hole entropy must be connected to the defining property of any horizon, i.e., observers on the one side of it cannot experimentally access the other side. Conversely, it needs to be insensitive to what precisely lies hidden behind the horizon. The most mainstream microscopic interpretation of black hole entropy consistent with these requirements is in terms of quantum entanglement of vacuum fluctuations of matter and gravitational fields across the horizon~\cite{Sorkin:1986,Srednicki:1993,Solodukhin:2011}. While the entanglement entropy interpretation is often tacitly assumed in the local equilibrium derivations, there exist other possibilities. In particular, diffeomorphism symmetries of a wide class of null surfaces (including event horizons and local causal horizons) form a Virasoro algebra with a central charge~\cite{Carlip:1999,Chakraborty:2016}. Then, the Cardy formula for the corresponding microstates yields the expected entropy for any causal horizon. Furthermore, even standard methods of calculating black hole entropy, e.g., the Wald prescription~\cite{Wald:1993,Wald:1994} and Euclidean partition function approach~\cite{Gibbons:1977} give the same result for both event and local causal horizons~\cite{Jacobson:2019,Jacobson:2022a,Jacobson:2022b}.

A potentially uncomfortable idea must be accepted: If we assign entropy to any causal horizon, it becomes an observer-dependent quantity. This is obvious, since an accelerating observer perceives a horizon and the corresponding entropy, whereas an inertial observer in the same place does not. However, due to the Unruh effect, the accelerated observer also detects a nonzero temperature. It has been shown that, in the presence of matter, one can use this temperature to compute the Clausius entropy flux the accelerated observer measures~\cite{Baccetti:2013ica} (whereas the inertial observer measures no such flux). Moreover, the density operators in quantum field theory are themselves observer-dependent~\cite{Gomez:2022}, and it follows that entanglement entropy computed from them has the same property. In view of this, we find the observer-dependence of entropy in relativistic physics to be a well-established, if not often stressed, fact rather than an additional assumption we must introduce.

The minimal assumptions necessary for local equilibrium conditions are thus the universality of the Unruh effect and of the expression for horizon entropy. One naturally asks what we get in return for embracing these assumptions. After all, both classical and semiclassical Einstein equations can be derived without any reference to local equilibrium. In this regard, one could think of the local equilibrium methods as a mere curiosity with little relevance for our understanding of gravity. One way to give them a deeper meaning lies in interpreting them as an indication that gravity emerges from statistical behavior of some unknown quantum degrees of freedom that make up the classical spacetime~\cite{Jacobson:1995ab,Padmanabhan:2010,Verlinde:2011}. Then, gravity would no longer be a fundamental interaction. Gravitons would acquire a similar status as phonons, being simply a convenient description of excitations of the underlying quantum structure of the spacetime. This interpretation is certainly possible and thought provoking. However, we can learn much from the local equilibrium conditions even without embracing it. We just require that the local equilibrium conditions \textit{encode} the gravitational dynamics, or at least its most important features. Thus, we treat them as a consistency condition that quantum gravity should recover, regardless of whether the gravitational force is ultimately emergent or fundamental. This is similar in spirit to the requirement that any viable theory of quantum gravity reproduces Bekenstein entropy in a suitable limit. Crucially, the local equilibrium approaches give us the ability to not only test the specific candidate theories, but to develop predictions valid regardless of the final theory of quantum gravity. In the following, we will discuss how the local equilibrium approaches applied to both classical and low energy quantum gravitational dynamics yield novel insights, not accessible by other known methods of studying gravity.

\section*{New insights into (semi)classical dynamics}

Following the previous discussion, let us now assume that suitable equilibrium conditions applied to locally constructed causal horizons encode \textit{all} the information about classical gravitational dynamics. Do we then find a dynamics equivalent to general relativity? Since one obtains equations of the same form as the Einstein field equations, the answer would seem to be affirmative. However, a closer look reveals that certain features of general relativity must be added on top of the local equilibrium conditions ``by hand''. In particular, local equilibrium conditions do not constrain the energy-momentum tensor to be divergence-free, in the way a fully diffeomorphism invariant action principle does. Instead, it only requires that its divergence is proportional to a gradient of an arbitrary scalar function~\cite{Alonso:2020a,Alonso:2021}. Nevertheless, suppose that we impose the condition that the divergence of the energy-momentum tensor vanishes. Then, integrating this condition yields an arbitrary integration constant, that can be identified with the cosmological constant $\Lambda$ in the Einstein equations. This differs from the status of $\Lambda$ in general relativity, where it appears as a fixed constant parameter in the  Einstein-Hilbert Lagrangian. To better appreciate the significance of this, consider the other fixed parameter present in the Lagrangian, the Newton gravitational constant, $G$. The local equilibrium approaches recover $G$ by relating it to the fixed proportionality constant in the horizon entropy~\cite{Jacobson:2016}. Hence, $G$ is a fixed parameter in both the Lagrangian and the local equilibrium viewpoint. The same applies to coupling constants of higher derivative corrections~\cite{Bueno:2017,Svesko:2017,Svesko:2019,Alonso:2023}. Therefore, $\Lambda$ stands out as the only fixed parameter in the gravitational Lagrangian that changes into an integration constant in the local equilibrium picture.

One may notice that both the possible nonzero divergence of the energy-momentum tensor, and the appearance of $\Lambda$ as an arbitrary integration constant are characteristics of Weyl transverse gravity~\cite{Alvarez:2006,Carballo:2022} (a particular formulation of unimodular gravity). This theory has the same classical solutions as general relativity, but its symmetry group differs. Rather than being invariant under all diffeomorphisms, the action of Weyl transverse gravity is invariant only under spacetime volume preserving (transverse) diffeomorphisms. Furthermore, it is invariant under Weyl transformations which multiply the metric by an arbitrary everywhere positive function and leave the matter fields unaffected. While we will not discuss technical details of Weyl transverse gravity, the theory has been extensively explored in a number of recent works, see, for example~\cite{Oda:2017,Barcelo:2018,Kugo:2021,Kugo:2022a,Kugo:2022b,Alonso:2022,Carballo:2022,Alvarez:2023,Garay:2023}. The upshot of these studies is that general relativity and Weyl transverse gravity have equivalent perturbative quantizations. Moreover, both theories appear to be equally compatible with various nonpertubative approaches to quantum gravity, e.g. string theory and Euclidean path integral formulation~\cite{Carballo:2022,Garay:2023}. There even exists a one-to-one correspondence between diffeomorphism invariant and Weyl transverse invariant modified theories of gravity, in the same sense in which general relativity corresponds to Weyl transverse gravity~\cite{Carballo:2022}.

The only so far uncovered physical difference between general relativity and Weyl transverse gravity is the aforementioned status of $\Lambda$. In general relativity, it is a fixed parameter in the Lagrangian, whereas it appears as an integration constant in Weyl transverse gravity. Stated differently, in the 3+1 formulation of Weyl transverse gravity we are free to specify an arbitrary initial value of $\Lambda$, whereas we have no such freedom in general relativity. This difference has important consequences for the so called cosmological constant problem. It is well known that the value of $\Lambda$ in general relativity acquires unrealistically large contribution due to quantum corrections~\cite{Weinberg:1989,Polchinski:2006,Burgess:2013}. Moreover, such high contributions arise from every further higher loop correction one adds, making $\Lambda$ radiatively unstable (in other words, an infinite number of counter-terms would be necessary to recover the value of $\Lambda$ observed in nature). In Weyl transverse gravity, $\Lambda$ turns out to be radiatively stable thanks to the breaking of the full diffeomorphism invariance~\cite{Carballo:2015}. This solves one of the two major problems related with $\Lambda$. The second one, i.e., finding the mechanism leading to its observed value, remains. Nevertheless, several proposals for its resolution within the context of Weyl transverse gravity have been put forward~\cite{Barcelo:2018,Perez:2018,Carballo:2020}.

The status of $\Lambda$ as an integration constant and the possible nonzero divergence of the energy-momentum tensor strongly suggest the local equilibrium approaches lead to Weyl transverse gravity~\footnote{An alternative interpretation of the behavior of $\Lambda$ in thermodynamics of spacetime has been put forward~\cite{Padmanabhan:2003,Padmanabhan:2014,Padmanabhan:2016,Padmanabhan:2022}. However, in this view, gravity is considered as an emergent phenomenon.}. However, to really trust this result, the local equilibrium conditions should also be able to recover the Weyl invariance. The first step in this direction is to show that, if we assume Weyl and transverse diffeomorphism invariance from the start, the local equilibrium conditions are equivalent to equations of motion of Weyl transverse gravity. That this is indeed the case can be already seen from the expression for Wald entropy in Weyl transverse gravity we derived\footnote{Furthermore, we explicitly showed that the local equilibrium conditions which are Weyl and transverse diffeomorphism invariant encode the equations of motion of Weyl transverse gravity~\cite{preparation}.} in reference~\cite{Alonso:2022}. However, this only proves that local equilibrium methods are compatible with Weyl transverse gravity. To show that they actually favor it over general relativity, we have to argue that the entropy of a local causal horizon is invariant under Weyl transformations, without appealing to the gravitational dynamics in any way. On the one side, for conformal quantum fields, it is straightforward to show that the matter entanglement entropy does not change under Weyl transformation (the situation with nonconformal fields is subtle~\cite{Jacobson:2016}). On the other side of the equality, we cannot perform such a simple check for the entropy of the horizon. However, the Unruh temperature of a certain realization of a local causal horizon, a causal diamond, can be derived by exploring an analogy with conformal quantum mechanics~\cite{Arzano:2020,Arzano:2021} (i.e., one dimensional conformal quantum field theory). If entropy can be computed in the same way, it will be completely independent of gravitational dynamics and explicitly Weyl invariant (recently, some work has been done towards defining entropy in this way~\cite{Gallaro:2022}). Then, it would follow that gravitational dynamics derived from Weyl invariant entropy is Weyl invariant as well, strengthening the link between the local equilibrium conditions and Weyl transverse gravity.

So, local equilibrium methods, when treated as our only source of information on gravitational dynamics, single out a theory of gravity that partially solves the cosmological constant problem. Can they also provide the second part of the solution, i.e., a suitable source for the small observed value of the cosmological constant? Possibly. Recently, a mechanism in which nonconformal quantum fields can source $\Lambda$ in Weyl transverse gravity has been proposed~\cite{Perez:2018}. The entanglement entropy of nonconformal fields contains a scalar contribution, $X$~\cite{Speranza:2016,Casini:2016,Longo:2020}. In local equilibrium approaches $X$ enters the gravitational equations as $Xg_{\mu\nu}$. Then, it is easy to see that $X$ effectively acts as a varying contribution to $\Lambda$~\cite{Jacobson:2016}. This term is typically suppressed in the final gravitational equations, since it is not compatible with full diffeomorphism invariance. However, in Weyl transverse gravity, we are free to keep it by relating it to the divergence of the energy-momentum tensor. For a massive field, $X$ has a contribution proportional to mass squared~\cite{Longo:2020}. In principle, this additional scale given by particle masses allows $X$ to act as a varying $\Lambda$ that is far below the Planck scale. Since the effective field theory Planck scale contributions to $\Lambda$ do not appear in Weyl transverse gravity, we would then have a consistent small value of $\Lambda$. Whether this value will be in agreement with the observed $\Lambda$ remains to be shown.

\section*{The logarithmic term and quantum corrections}

We have argued that the local equilibrium methods provide new insights already at the level of (semi)classical gravitational dynamics. We are now interested whether they also provide hints about quantum corrections beyond the semiclassical Einstein equations. On the one hand, the equations of motion (both classical and semiclassical) for a wide class of modified theories of gravity were derived from local equilibrium conditions~\cite{Eling:2006,Padmanabhan:2010,Jacobson:2012,Bueno:2017,Svesko:2017,Svesko:2019}. On the other hand, these derivations rely on the Wald prescription for entropy. Since Wald entropy is obtained directly from the gravitational Lagrangian, using it in a way presupposes the answer. One actually shows that the Noether charges of diffeomorphism invariant (or transverse diffeomorphism invariant) theories of gravity contain sufficient information to reproduce their equations of motion. While interesting in its own right, this observation does not really constrain our search for quantum gravity in any way.

We instead want to study the quantum corrections without committing to any particular theory. Ideally, we should find an entropy prescription that explicitly includes higher order quantum corrections while being independent of any particular realization of quantum gravity. Then, gravitational dynamics derived from such a modified entropy prescription should include quantum corrections valid regardless of the final theory of quantum gravity. Any results obtained in this way will of course be limited to the regime in which the spacetime is still describable as a Lorentzian manifold (we will mention a possible generalization later on). However, they can still suffice to provide hints about the nature of quantum gravity and, in principle, even predict observable effects.

Remarkably, most approaches indeed agree on the same modification for the entropy of a causal horizon. They predict that, in four spacetime dimensions, it has the structure
\begin{equation}
S=k_{\text{B}}\mathcal{A}/\left(4l_{\text{P}}^2\right)+k_{\text{B}}\mathcal{C}\ln\frac{\mathcal{A}}{\mathcal{A}_0}+O\left(\frac{\mathcal{A}_0}{\mathcal{A}}\right),
\end{equation}
where $\mathcal{A}$ is the area of a suitable spatial cross-section of the horizon, $\mathcal{A}_0$ is a constant with dimensions of area (usually taken to be of the order of $l_{\text{P}}^2$) and $\mathcal{C}$ is a number, whose sign and value are theory-specific. This form of entropy appears, e.g. from string theory~\cite{Banerjee:2011,Sen:2013,Karan:2023}, loop quantum gravity~\cite{Kaul:2000,Meissner:2004}, AdS/CFT correspondence~\cite{Faulkner:2013},  entanglement entropy calculations~\cite{Solodukhin:2010,Solodukhin:2011}, generalised uncertainty principle phenomenology~\cite{Adler:2001,Medved:2004}, quantization of the horizon area~\cite{Hod:2004,Davidson:2019}, nonlocal effective field theory~\cite{Xiao:2021}, the existence of minimal resolvable area~\cite{Alonso:2021b}, and from model independent analysis of statistical fluctuations~\cite{Gour:2003,Medved:2005}.

Aside from its widespread appearance, the logarithmic term has further attractive properties. First, the coefficient $\mathcal{C}$ is dimensionless and, within a specific theory, it tends to be a universal constant. As an example, for entanglement entropy $\mathcal{C}$ is fully determined by the conformal anomaly and can be precisely computed, once the matter content of the theory is specified~\cite{Solodukhin:2011}. In contrast, the leading order term proportional to area depends on an undetermined UV cutoff. Similarly, loop quantum gravity leads to the universal value $\mathcal{C}=-3/2$, whereas the leading order term in entropy involves the Barbero-Immirzi parameter, which is in principle arbitrary (and usually fixed by the requirement of recovering the correct Bekenstein entropy). Second, in the limit of $l_{\text{P}}\to0$, the leading order and logarithmic term are the only divergent ones, whereas the $O\left(\mathcal{A}_0/\mathcal{A}\right)$ subleading terms vanish. This suggests that the logarithm might significantly affect the low energy quantum gravitational dynamics (i.e., in the regime in which $l_{\text{P}}$ is not negligible, but still significantly smaller than other relevant length scales), whereas the $O\left(\mathcal{A}_0/\mathcal{A}\right)$ terms are unlikely to do so.

To sum up, we have a strong motivation to look at the implications of the logarithmic term for gravitational dynamics. So, what happens if we include it in the local equilibrium conditions? To gain some intuition, we can first derive the linearized equations for gravitational dynamics. In this case, with some technical caveats, we obtained the linearized equations of motion of quadratic gravity (or, more precisely, of its Weyl transverse formulation), with the higher derivative terms suppressed by factor $\mathcal{C}l_{\text{P}}^2$~\cite{Alonso:2023}. Since quadratic gravity is the unique purely metric, (transverse) diffeomorphism invariant theory of gravity in four dimensions, this result is of course expected. Thus, the recovery of linearized quadratic gravity serves as an important consistency check. It shows that the relation between gravitational dynamics and the local equilibrium conditions continues to hold even if we include the low energy quantum gravitational effects.

It could be expected that going beyond the linearized regime simply leads to the full quadratic gravity. However, this is not the case. We carried out an analysis of the nonlinear regime under certain simplifying assumptions, uncovering a correction proportional to $\mathcal{C}l_{\text{P}}^2R_{\mu\rho}R_{\nu}^{\;\:\rho}$~\cite{Alonso:2020b}. Moreover, the behavior of local causal horizons under fluctuations of the metric in the absence of matter suggests that corrections proportional to the Weyl tensor squared should appear as well~\cite{Jacobson:2017,Wang:2019}. No such terms are present in the equations of motion of quadratic gravity. In fact, there exist no purely metric, transverse diffeomorphism invariant theory of gravity in four-dimensional spacetime which can contain them. Then, to interpret the low energy quantum gravity corrections, we have to either break the transverse diffeomorphism invariance or to include additional fields. While we do not yet know which answer will prove to be correct, we have a strong clue. The equations of motion of the so-called four dimensional Einstein-Gauss-Bonnet gravity include the same ``Ricci squared'' term we found~\cite{Glavan:2020}.  While this theory unfortunately leads to a number of pathologies~\cite{Gurses:2020,Ai:2020,Shu:2020,Arrechea:2021}, these are all cured by its scalar-tensor generalization~\cite{Lu:2020,Hennigar:2020}. This formulation of Einstein-Gauss-Bonnet is fully diffeomorphism invariant and well-defined in four (or even fewer) spacetime dimensions. Hence, it is tempting to speculate that local equilibrium conditions encode (Weyl transverse invariant) gravitational dynamics that combines the scalar-tensor Einstein-Gauss-Bonnet theory and quadratic gravity. Of course, to show this, we will have to identify an additional scalar degree of freedom (besides the metric) in the local equilibrium conditions. Regardless of its precise form, the final result will be modified gravitational equations for low energy quantum gravitational dynamics with correction terms suppressed by $\mathcal{C}l_{\text{P}}^2$. The dependence of the modified equations on the underlying theory of quantum gravity will be completely captured by the value and sign of the coefficient $\mathcal{C}$.

\section*{Summary and future perspectives}

In summary, local equilibrium conditions are intimately related with gravity. If we accept that they encode all the information about gravitational dynamics, they seem to imply the following picture: i) The symmetry group of gravity consists of transverse diffeomorphisms and Weyl transformations; ii) Gravitational equations contain quantum corrections which combine terms corresponding to quadratic gravity and to scalar-tensor Einstein-Gauss-Bonnet gravity, all suppressed by a factor $l_{\text{P}}^2$; iii) Local equilibrium approaches may provide an explanation for the observed cosmological acceleration, relating the value of $\Lambda$ to the presence of nonconformal quantum fields. Of course, the details of the quantum modified gravitational dynamics are yet to be understood and the suggested explanation for $\Lambda$ is currently just an educated guess. Nevertheless, we obtain a compelling and fairly conservative framework, implied just by imposing some reasonable requirements on the entropy of local causal horizons. Altogether, this framework both provides theoretical hints for the final theory of quantum gravity, and has phenomenological consequences, e.g. for the early universe physics (which we already studied in a simplified setting~\cite{Alonso:2023b}) and the structure of black holes.

To conclude, we briefly outline a possible way to check our ideas from a different angle, and possibly bypass some of the limitations of the local equilibrium approaches. A procedure to reconstruct the properties of spacetime (and even gravitational dynamics~\cite{Perche2}) by performing quantum measurements was recently proposed~\cite{Perche:2022}. The measurements recover the Wightman function, whose derivatives then encode the metric and the Riemann tensor\footnote{A similar approach to metric and curvature reconstruction has been previously put forward in a different context~\cite{Kothawala:2013,Kothawala:2014}.}. The Wightman function has the same structure of UV divergences as the modified Bekenstein entropy with a logarithmic correction. Hence, our approach to derive quantum corrections to gravitational dynamics from the logarithmic term should work here as well. Notably, the quantum measurements approach does not rely on a notion of thermodynamic/entanglement equilibrium. Hence, it might in principle allow us to study higher energy quantum gravitational regimes.

\end{document}